# Radioplasmonics: design of plasmonic milli-particles in air and absorbing media for antenna communication and human-body in-vivo applications.

Ricardo Martín Abraham-Ekeroth*†,‡,¶

†Istituto Italiano di Tecnologia (IIT), Via Morego 30 16163 Genova, Italy

‡Instituto de Física Arroyo Seco, IFAS (UNCPBA), Tandil, Argentina

¶CIFICEN (UNCPBA-CICPBA-CONICET), Grupo de Plasmas Densos, Pinto 399, 7000 Tandil, Argentina

**ABSTRACT**

Surface plasmons with MHz-GHz energies are predicted by using *milliparticles* made of metamaterials that behave like metals in the radiofrequency range. In this work, the so-called *Radioplasmonics* is exploited to design scatterers embedded in different realistic media with tunable absorption or scattering properties. High-quality scattering/absorption based on plasmon excitation is demonstrated through a few simple examples, useful to build antennas with better performance than conventional ones. Systems embedded in absorbing media as saline solutions or biological tissues are also considered to improve biomedical applications and contribute with real-time, in-vivo monitoring tools in body tissues. In this regard, any possible implementation is criticized by calculating the radiofrequency heating with full thermal simulations. As proof of the versatility offered by radioplasmonic systems, plasmon "hybridization" is used to enhance near-fields to unprecedented values or to tune resonances as in optical spectra, minimizing the heating effects. Finally, a monitorable drug-delivery in human tissue is illustrated with a hypothetical example.

This study has remarkable consequences on the conception of plasmonics at macroscales. The recently-developed concept of "spoof" plasmons achieved by complicated structures is simplified in Radioplasmonics since bulk materials with elemental geometries are considered.

**INTRODUCTION**

Conceived originally in optics, *Plasmonics* has been evolving due to the increasing possibilities offered by nano-technology, extending its domain to higher and lower energies. Its main relevance is attributed to its sensitivity and versatility at small scales compared with its simple working principles [1]. In a typical metal nanoparticle, electrons of the conduction band oscillate with a characteristic resonance frequency that depends on the geometric and constitutive conditions, defining the localized surface plasmon. This sub-wavelength excitation can focus light at the nanoscale as well as "turn on" molecules [2], sense and heat small-scale biological environments [3,4], enhance Raman signals [5–7], emit or receive radiation as antennas [8,9], inhibit or kill viruses [10], favor chemical reactions by providing "hot spots" [11], among other fascinating phenomena.

Simultaneously, the appropriate manipulation of methods, materials, and techniques in meso and nanoscales facilitated the development of *Metamaterials* (MMs) [12]. Artificially-made, this kind of material is generally structured in a sub-wavelength scale by composing several materials or by altering the geometry such that the whole material may achieve a desired, sometimes unusual electromagnetic (EM) property [13,14]. In this way, much of the effort in MMs' progress has been dedicated to the total control of EM fields, as a kind of generalization of plasmonics and other branches of photonics, independent of the photon energy, materials, and the scales involved [15,16].

In particular, the control of the EM field in the radiofrequency (RF) range has become essential for the increasing antenna technologies, information networks, and Internet of Things [17]. In this regard, more than one century of experience accompanies the development of RF devices since the invention of the antenna [18,19]. However, modern technologies like Radio Frequency Identification (RFID) or Near-Field Communication (NFC) demand continuous device efficiency and miniaturization [17,20]. Simplest designs would also be beneficial. Recently, a few miniature devices based on the so-called *spoof plasmons* (SPs) have been fabricated to work as RF antennas [21–24]. These devices can be considered as *plasmonic metamaterials* [25] that use altered geometries to achieve an average plasmonic behavior in the RF domain. However, they in general involve complex designs and are poorly versatile since they are built upon a fixed geometry [17,24,26,27]. Complex designs can lead to a difficult comprehension of the phenomena involved and a difficult comparison with the knowledge in conventional antennas [19].

Moreover, from a theoretical point of view, the complex light-matter interactions that occur in plasmonic nanosystems need to be better understood for future developments in light manipulation. A bigger-scale framework of these interactions would be significant provided that easy and more controllable experiments could be performed with lower energies and costs. As an example, the plasmonic nanoantennas have been always compared with their RF counterparts [19]. However, there are several differences regarding size (sub-wavelength vs. wavelength-scale devices), the skin depth in the materials involved, and the behavior of their dielectric/permittivity functions [28]. This contrast has made the knowledge in RF implementations laborious to apply to the nanoscale equivalents.

On the other hand, a lot of research has been done to apply plasmonics in biomedicine [29–31], using the advances in this latter field in combination with nanotechnology. Nevertheless, the use of optical and infrared radiation cannot penetrate much into real tissues so the inherent exploration possibilities cannot be exploited. In contrast, the use of RF radiation may offer a solution since tissues are greatly penetrable at this range [32–34]. Yet, the RF treatment is usually not localized and difficult to control, thus adding potential damages due to thermal and undesired EM effects [32,35–37]. Moreover, some fundamental applications like the RF treatment of tumor ablation, drug delivery, and *in-vivo* monitoring of tissues can be invasive and painful [32,37]. Other applications in both air and water-like surroundings may need high electric fields in a short distance to work appropriately as food treatments, sterilization, etc. [38,39].

In a recent work by the author [34], modern metamaterial (MM)-based mesostructures were highlighted with a versatility that is unprecedented by conventional nanomaterials, especially in the radiofrequency (RF) domain [40–47]. These so-called *RF-metamaterials* (RF-MMs) are obtained by "doping" polymers or dielectric matrices with carbon-based nanomaterials or nanostructured metals. In this way, the electrical conductivity can be modified in the whole resultant material to switch the isolator/conductor condition [47]. The MM is designed to overcome the *percolation threshold*, and metallic-like conduction is obtained in the composite. Specifically, negative permittivity is achieved and plasma-like behavior can be found in a way similar to that occurring in metals for the optical regime. In other words, this new class of hybrid materials, that behave like metals, could pave the way to create new RF devices and small underwater systems for in-vivo biological application, provided that they can beat the limit of deep-tissue penetration.

In particular, Ref. [34] defined a new research topic called Radioplasmonics: a resonant absorption using RF-MM microparticles was achieved in RF by surface plasmon excitation. This phenomenon was applied to design micro-transducers that improve photoacoustic signals.

In this paper, the concept of Radioplasmonics is focused on scattering phenomena in RF in addition to absorption to develop new applications. A novel Plasmonic concept is shown in macroscale with particle dimensions less than a few millimeters or centimeters [48], where it would be easy to carry out experiments to study complex plasmonic phenomena as, for instance, multiple particle coupling [21]. This work naturally includes previous efforts to define plasmonics in extremely low-energies [27,48–52]. The latest ideas about SPs are simplified since "bulk" materials with simple geometries are used here. The complexity of device design based on SPs is avoided, provided that the MM's working principles are subjacent in their mesoscopic structure [25]. The study takes profit of the background accumulated in optical plasmonics to exploit plasmon resonance tuning, sensitiveness to surrounding media, near-field amplification and confinement or, more generally, the control based on the geometry and constitution of the system [2,28,53]. Moreover, the efficiency of *radioplasmonic* (RP) antennas is proven to be better than conventional RF-antennas. The present ideas about RP-antennas are fundamental in any modern telecommunication device working in the GHz range, regarding both efficiency and miniaturization. Moreover, both huge field confinement and amplification are shown to be possible in near-field; here confinement orders as $\sim \lambda/10^6$ (usual plasmonics deals with $\sim \lambda/100$) and enhancements as high as 3500 times the field of an incident wave are obtained. Besides, scatterers in absorbing embedding media as saline solutions or real tissues are also studied [36,54]. Miniature antennas are designed to work in MHz range to facilitate radiation penetration in tissue. This numerical approach may benefit non-invasive biomedical applications and pay the way for a new real-time, *in-vivo* monitoring tool in body tissues [55] since high-penetration of tissues and small sizes are taken as an advantage. Moreover, any possible realistic implementation is criticized by analyzing RF heating [36]. Finally, a drug-delivery application inside a real human tissue is illustrated by the use of an RP particle.

**METHODS**

Differently from the previous work in Radioplasmonics [34], full EM solutions are performed that work beyond the quasistatic approximation [2], provided that the bigger particles simulated here cannot be represented, in general, by single dipole moments. A Mie code is implemented for bare spherical particles immersed in absorbing media by following the works [56,57]. This study is also supported by Mie calculations for core-shell spherical structures based on the Refs. [58,59] that correspond to a particle embedded in non-absorbing media. As a complement and to compare the results, COMSOL Multiphysics software is used; the FEM method is applied to calculate the EM response by several particle shapes in absorbing media. Finally, COMSOL software is also used to fully simulate the RF heating of the system particle+host medium under continuous wave (CW) radiation, taking into account the maximum RF-field allowed by safety regulations stated in Ref. [35].

**RESULTS AND DISCUSSION**

To start the discussion about RP excitation, the EM response by several RF-MM particles is shown (Figure 1). The aim is also to demonstrate the feasibility of the phenomenon by using realistic composites recently reported. The particles consist of spheres of diameter 1 cm embedded in air. Figure 1a (b) shows the scattering (absorption) efficiency as a function of the frequency in the GHz range. The composites used correspond, respectively, to: graphene dissolved at 3 %wt. in a Polydimethylsiloxane (PDMS) matrix [46] (black solid line), a solution of Al in Acrylic/Polyurethane at 86 %vol. [47] (red line with squares, APu/Al), multi-walled carbon nanotubes (MWCNT) in Alumina at 12 %wt. [43] (green line with circles), carbon nanofiber (CNF) at 6 %wt. embedded in silicone [42] (blue line with up-pointing triangles), a nickel(II) oxide/titanium nitride (NiO/TiN) composite[60] with its ratio $x$ of TiN and NiO equal to $x = 0.6$ (violet line with down-pointing triangles), an amorphous alloy of $Fe_{78}Si_9B_{13}$ dispersed in an epoxy matrix with a volume fraction of 78 %vol. [40] (brown line with diamonds), a Polypyrrole (PPy) composite [61] having a concentration of carbon nanotubes (CNTs) of 9% wt. (magenta line with left-pointing triangles), and finally the previous solution of NiO/TiN composite with a ratio $x = 1$ (orange line with right-pointing triangles).

Note that several peaks appear in the responses that are easily identifiable with plasmonic excitations. Mie theory is used to identify the multipolar plasmon resonances as it is usually performed for plasmonic examples in the optical range [2]. The interested reader can see another example given in the Supplementary Information (SI) for a bigger particle where a multipole analysis and near-field patterns are shown. It is clear from Figure 1 that each MM contributes with a different spectrum provided that their effective dielectric functions are different. These materials have been chosen to show their main resonances at different frequencies for distinct purposes. Care must be taken if the multipolar orders interfere with the purposes. As in optics, this effect can be avoided by changing the size, the particle shape, and/or the embedding material. These examples are also thought to be useful for studying plasmonics at big scales and they may allow for easy experiments with cheap and traditional RF devices.

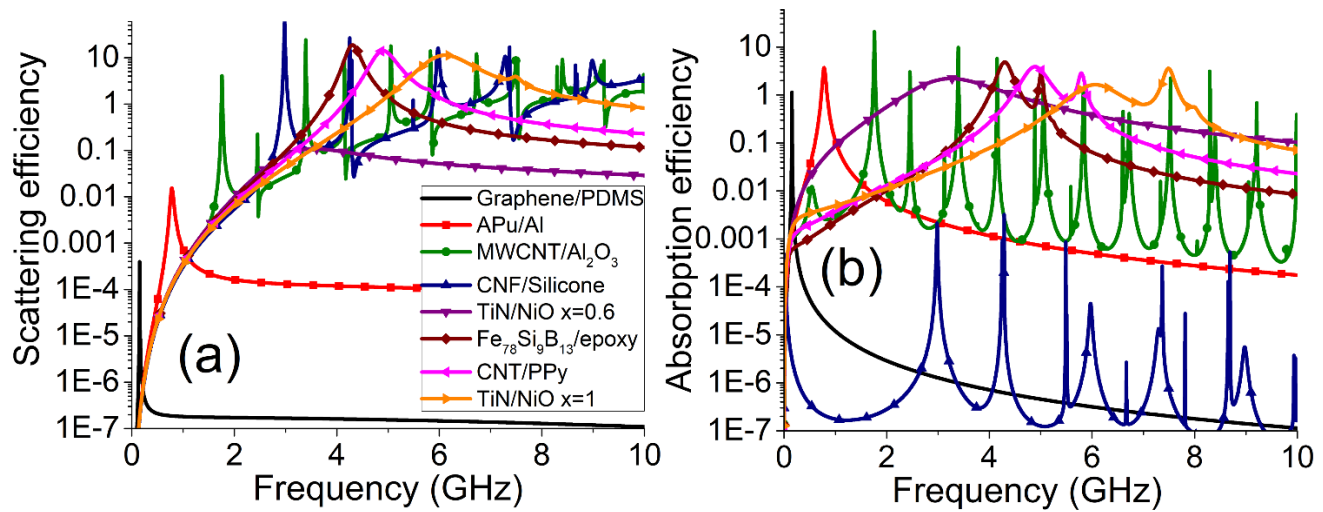

Figure 1: Electromagnetic efficiencies for spherical particles of diameter 10 mm as a function of the RF-metamaterial; a) scattering, b) absorption.

Although simple, the spherical particles may not constitute the most suitable shapes for antenna design, nor they might be easy to fabricate. To illustrate the shape dependence of Radioplasmonics, the EM response by several RP rods is shown in Figure 2. Both far (a-b) and near fields (c) are analyzed. The scattering performance of the rods in the far-field is also evaluated for antenna construction. These rods resemble wires in a typical RF antenna; they have their extremes rounded as the ones given in usual optical nanorods, thus this study serves also for direct contrast to nanorods' results. The rods are assumed to be made of PPy/CNT 9 %wt. as simulated in the example in Figure 1 (magenta line with left-pointing triangles). It is known that in the case of rods, two types of surface radioplasmons (SRPs) exist that are related to the two principal directions of the scatterer and the polarization state of the incoming wave, namely the long and short-axis SRPs. The direction and polarization of the incident plane wave are given in the scheme in Figure 2b, where $\boldsymbol{E}$ is the electric field and $\boldsymbol{k}$ is the wavevector.

Figure 2a shows several scattering spectra corresponding to a variation in the aspect ratio, $\chi$, given a fixed rod diameter of 7.4 mm. $\chi$ is defined as length/width (or rod's diameter) and their values give lengths that range from 7.4 mm up to 7 cm.

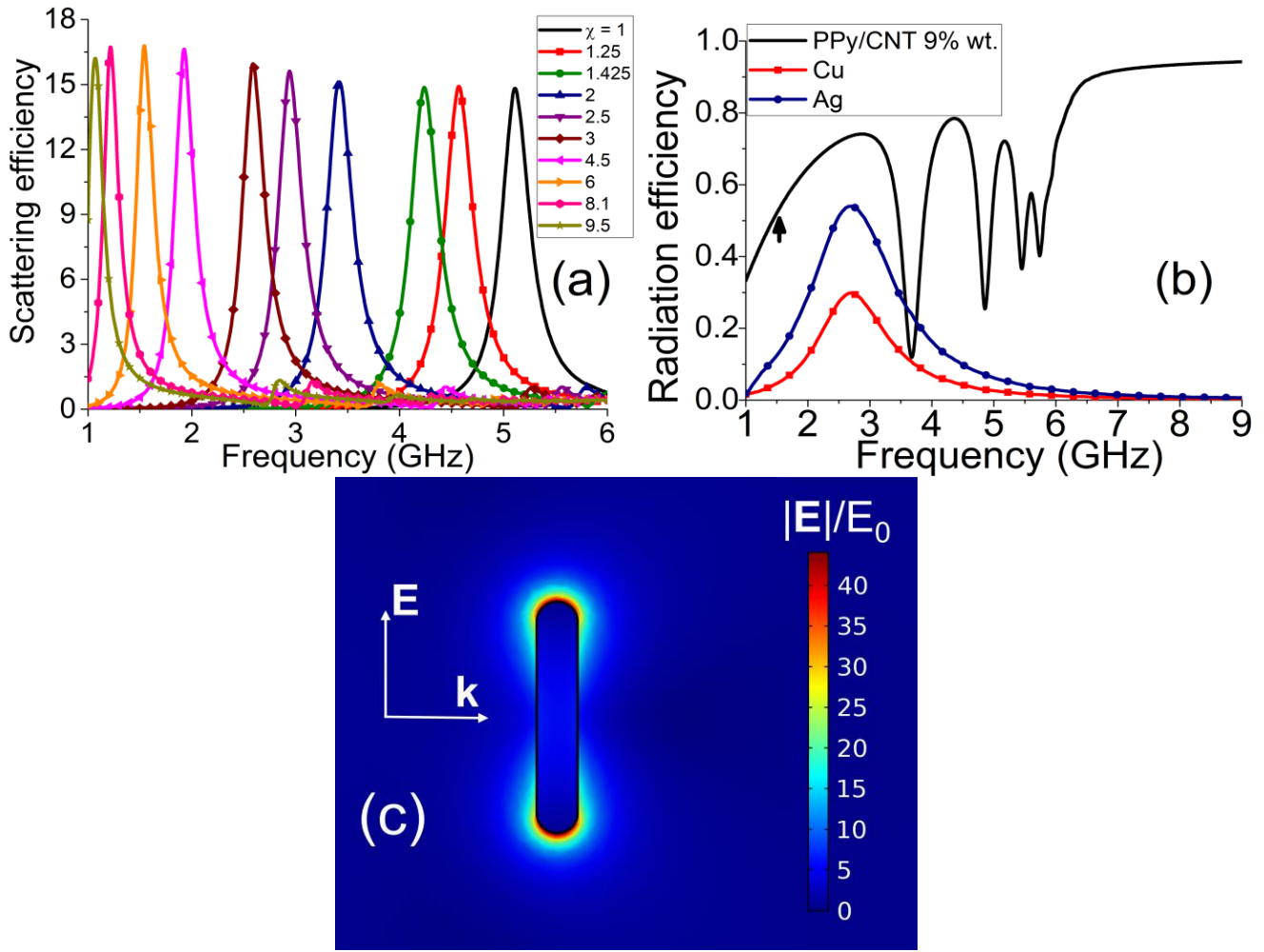

Figure 2: Electromagnetic response for radioplasmonic rods of PPy/CNT 9%wt in air having widths of 7.4 mm; a) scattering efficiency as a function of $\chi$, the aspect ratio. b) radiation efficiency for case $\chi = 6$ as a function of the material used. The arrow points to the location of the main RP resonance. c) near-field distribution of the scaled electric field for case $\chi = 6$ at 1.55 GHz.

Figure 2a demonstrates that a wide tuning of SRPs can be obtained to work in several bandwidths, useful for different applications. In this particular example, the resonances' spectral positions corresponding to the excitation of the dipolar, long-axis SRPs range from 1.1 GHz up to 5.1 GHz. However, thanks to the plasmonics characteristics of these devices, namely dependence on the size, shape, and environment of the SRPs, the resonances are easy to tune and the scattering intensities easy to control. For instance, the case $\chi = 6$ has a bandwidth lying around the GPS band for mobile phones, i.e. 1.575 GHz. The typical bandwidths of Bluetooth, Wi-Fi, and LTE systems of mobile communication can be easily covered and the system can be optimized to have a high-quality peak at the appropriate spectral location. Besides, more SRPs of higher orders are being excited in the structure. These appear in Figure 2a as the small peaks in the curves for frequencies higher than the main, dipolar resonances in each case.

As a complement of the study realized in Figure 2a, the interested reader can find in the SI the spectra of the corresponding absorption efficiencies for all the examples shown here.

Figure 2b shows the performance of the rod with $\chi = 6$ as a scatterer compared with typical rods used in RF antennas. The radiation efficiency is defined as equal to the ratio between the scattering and the extinction efficiencies, in a form similar to the one defined in antenna design [18,19]. Being the best available conductors for fabrication, copper and silver are used in the comparison. Remarkably, the efficiency for the RP rod is the best in almost all the frequencies calculated. The best *relative* performance is found at the spectral location of the main (dipolar) resonance, highlighted with an arrow. The rest of the multipolar contributions appear in Figure 2b as valleys or dips. Figure 2b means that the scattering cross section is significant, provided that the excited SRPs contribute also to the absorption cross section.

Note that the efficiencies by metal rods (lines with symbols) have a single peak in the range calculated, being its spectral location the same in both cases. Its frequency corresponds to the resonance occurring at $\frac{\lambda}{2} = length$ as in usual rods used in half-wave dipole antennas. Differently, the dipole resonance in the RP rod follows a condition $\frac{\lambda_{eff}}{2} \sim length$ where $\lambda_{eff}$ is an effective wavelength that depends on the material and geometric properties of the system [62]. This rule is equivalent to the well-known rule $\frac{\lambda_{SRP}}{4} \sim length$ in the half-wave dipole nanoantenna design [19,28,62], where $\lambda_{SRP}$ is the air/vacuum wavelength for the long-axis dipolar SRP. Similar to nanoantennas' theory, the difference between a traditional RF and an RP antenna lies in the contrast between materials: a good/perfect conductor is not equal to a plasmonic material since the penetration depth compared to the rod's size is different.

Another interesting feature to examine regarding radioplasmonics is the near-field distributions. Figure 2c shows a map of the electric field around the rod for case $\chi = 6$ at the resonance frequency, i.e. 1.55 GHz (compare to the orange line with right-pointing triangles in Figure 2a). This map shows the excitation of the long-axis SRP with a high field enhancement, resulting in more than 40 times the incident field. The field confinement is about $\lambda_{SRP}/4.4$, but given that the field spots are confined to the rod's widths [19], indeed the real confinement of the field reaches a ratio of $\lambda_{SRP}/26.3$ for the dipole excitation.

It is not the scope of this paper to provide details to build RF antennas nor compete against complex antenna designs as the typical patch antennas or PIFA of a cellphone, for instance. However, the phenomena lying in Figure 2 may open a new route to optimize antennas based on the radioplasmonics' concept. Several combinations of RF-MMs and shapes could contribute to building the next generation of antennas to cover 5G networks or beyond the current communication trends. Also, several SRPs of the same/different multipolar order/s can add up or hybridize when using more complex shapes [53,63] as seen below in this work.

Now the focus is moved to RP scatterers in water as embedding medium. The refractive index of water is known to be very high in the RF range [64]. This fact is used in combination with Radioplasmonics to miniaturize the scatterer and make it more useful for biomedical applications as follows in the paper. In the previous paper [34], the research was dedicated to studying very small particles with non-negligible thermal expansion to exploit photoacoustic applications given that all the EM energy reaching the particle is absorbed with negligible scattering (i.e. quasistatic approximation). Instead, bigger scales are taken here to make the scattering process also available in underwater systems.

The presence of ions diluted in water remarkably changes its EM properties by allowing water to strongly absorb the RF radiation. In this regard, it is common to assume a strong dependence of the refractive index on the salt type and concentration as well as on the temperature of the solution [65–67]. Here, Strogryn's model [65] is assumed to compare the EM properties of absorbing water with those for distilled (DI) water for later purposes. In what follows, a maximum size of only 3.6 mm is considered. A convenient RF-MM is used for RP scatterers, namely a glass composed with $Bi_2O_3$ and $SiO_2$ with a concentration of MWCNT of 20 %wt. which is abbreviated as BS/MWCNT glass 20 %wt. as reported in Ref. [68]. The scattering (absorption) efficiencies for a spherical particle are presented in Figure 3a (b). In both figures, DI and absorbing water as embedding media are compared.

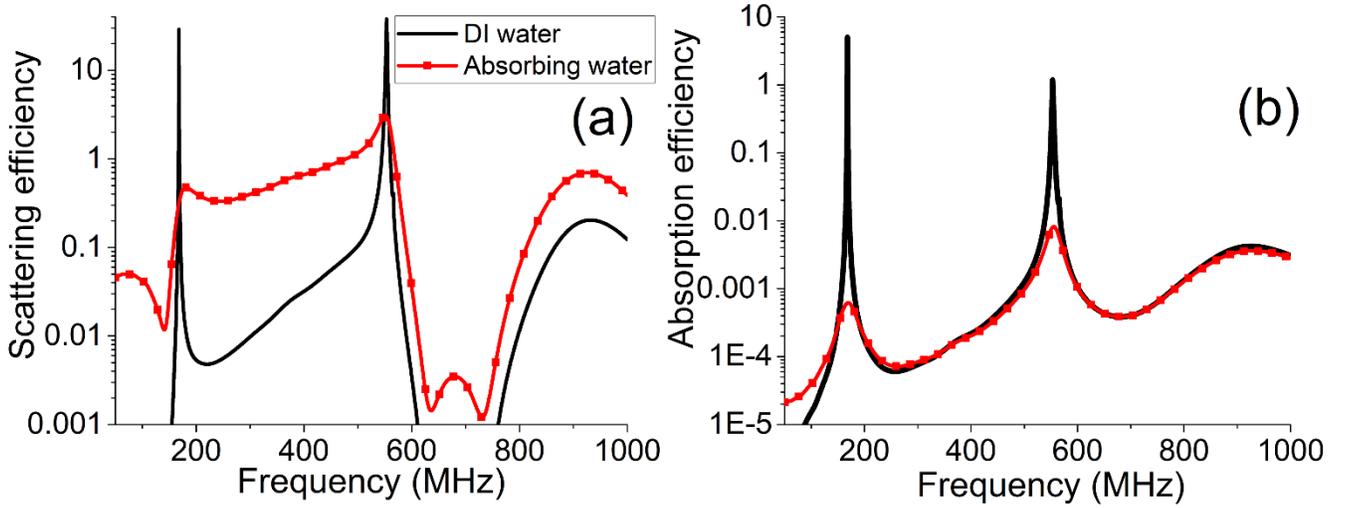

Figure 3: Electromagnetic efficiencies for a spherical particle of BS/MWCNT glass 20%wt. in water. Size is 3.6 mm. Black solid line (Red line with squares) corresponds to DI (absorbing) water. The absorbing water consists of a $Na^+$ + $Cl^-$ solution with salinity 3.2 ppt that simulates a usual body fluid; a) scattering, b) absorption.

For the case of absorbing water, a simplified NaCl solution is assumed (red line with squares in Figure 3, only $Na^+$ and $Cl^-$ ions). A salinity value of 3.2 ppt is simulated that corresponds to the typical body-fluid concentration [69]. A moderate scattering response is obtained due to the RP excitations in a suitable RF range where typical penetration depths lie in the meter scale in several body tissues [33], Figure 3a. Unfortunately, the effect of absorption of water is important, strongly decreasing the scattering efficiency. However, the absorption efficiency is also decreased with the presence of ions in water, Figure 3b. This is due to the fact that the absorbing water prevents the incoming radiation to reach the particle (*screening* effect), thus lowering the efficiency of the RP excitations in the particle surface.

Although a lower frequency range should be even more suitable for tissue applications, as biological entities are more transparent to lower energies, the signal bandwidths would be narrowest thus preventing antennas' interchange of information with high data rate. Then the range studied here can still be useful to conceive real applications.

Below, the EM properties of RP systems in saline solutions are analyzed to evaluate in-vivo and real-time body theragnosis. One of the most important aspects to be considered in this regard is the EM heating of tissues and saline solutions [32]. This important aspect can be taken as a plus for applications as aesthetic medicine [37] or photothermal therapy, including RF hyperthermia [32,70,71]. However, safety levels of irradiation need to be evaluated to avoid tissue damage. In what follows, electrolyte solutions are simulated as embedding media to evaluate their RF heating, provided also the importance of electrolyte solutions in tumor ablation applications and their contrast with tissues for imaging [32,36].

The RF heating of electrolyte solutions was analyzed recently in Ref. [36]. Simple formulas were deduced for the average absorbed power density $P$, namely equations 7-9 of that work. This formulation can be expressed in terms of the conductivity $\sigma$ of the solution and its real part of the relative dielectric function, $\varepsilon_r$, as

$$P(\omega) = \frac{1}{2}\sigma E_0^2 \frac{\omega^2}{\omega^2 + \left(\frac{\sigma}{\varepsilon_r \varepsilon_0}\right)^2}. \qquad (1)$$

This expression means that a bell-shape function can be obtained for the absorbed power as a function of $\sigma$ for a given frequency. This phenomenon corresponds to the heat generation in the solution due to the ions' oscillations and depends macroscopically on the EM properties of the solution. In other words, the model assumes that the viscous friction of the oscillating ions is the cause of the RF heating in the solution. As reported in [36], this model can be applied to NaCl solutions because it is independent of the type of ions involved. Besides, in Strogryn's model, the conductivity of the absorbing solution depends on the salinity or electrolyte concentration in water. This fact is taken into account in the present study to explore the heating rate of the solution itself, without any particle, as a function of its salinity (Figure 4a). Note that different peaks are obtained as a function of the excitation frequencies calculated, i.e. 168 and 555 MHz (see scattering peaks in Figure 3a).

Noteworthy, this peak-dependence on the frequency plays a key role in the evaluation of the RF heating in realistic water solutions if one intends to use RP phenomena, and it is very different from what can be found in optical photothermal therapy when nanoparticles are used. The contribution due to equation (1) is negligible in the optical range.

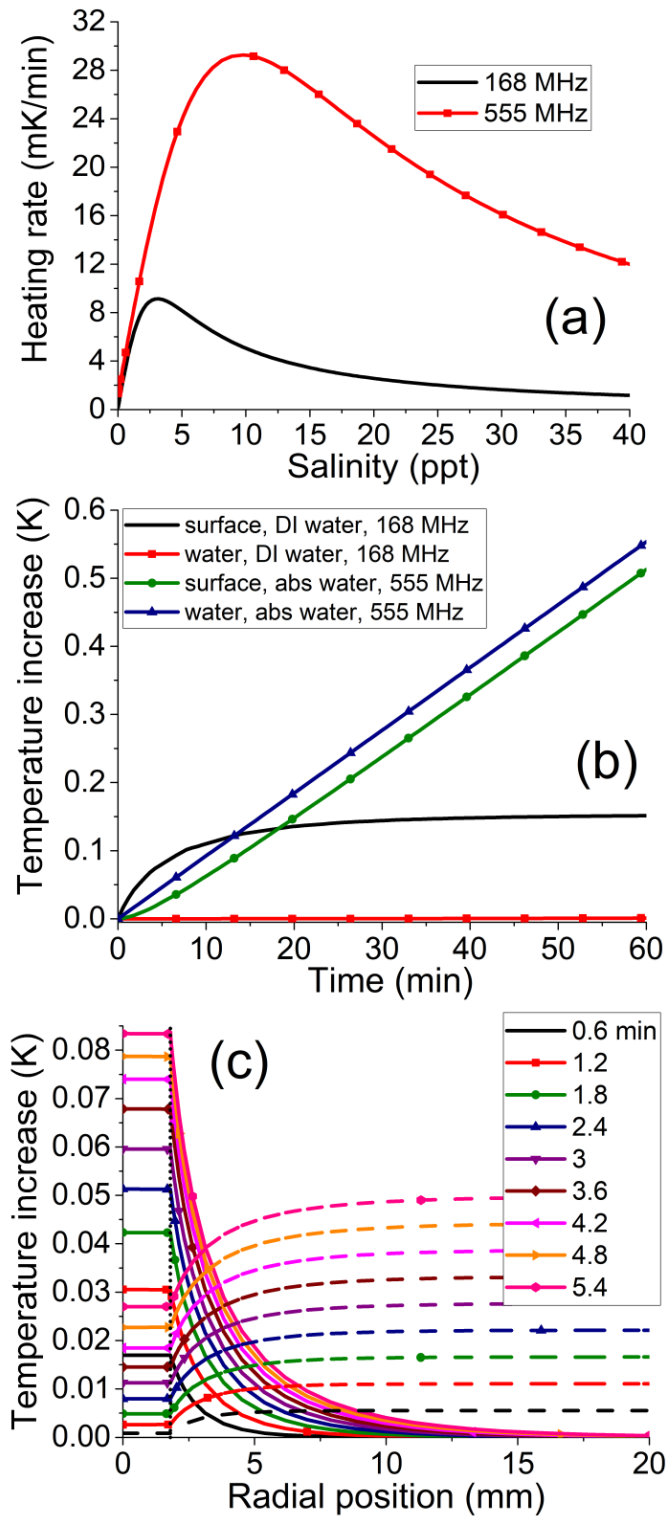

Figure 4: RF heating for systems in water-like media under CW radiation. An electric-field amplitude of 61 V/m is used in agreement with the maximum allowed value for safety in human tissues. (a) Heating rate of water as a function of its salinity (NaCl) for the values of SRPs resonances seen in Figure 3. (b-c) Temperature-increase for the system particle+water of Figure 3 at the SRPs resonances; the salinity is 3.2 ppt. (b) Time-evolution, temperature values at the particle surface are compared against values in water far away using both DI and absorbing water as in Figure 3; (c) Profiles of the temperature-increase as a function of radial distance for several values of time. Solid (dashed) lines correspond to DI (absorbing)

water at 168 (555) MHz. These frequencies correspond respectively to the maximum absorption obtained for each case, see Figure 3b.

The frequencies in Figure 4a correspond to the SRPs resonances in case of particle+absorbing water, red line with squares in Figure 3a. Although the heating rates do not reach their maxima for a salinity value of 3.2 ppt as in the example of Figure 3a, it is clear from the curves that a remarkable temperature increase must be obtained due to the effect described in Eq. 1. In the case of the curve for 168 MHz, the maximum is very close to the typical salinity value of 3.2 ppt for body tissues. Moreover, and as expected, the higher the excitation frequency the higher the amount of heat released by minute [36]. Thus, the results suggest that absorbing-water heating should be carefully considered when typical experiments with saline solutions are performed.

In Figure 4b and 4c, the temperature increase is compared for the system BS/MWCNT glass 20%wt. particle+water presented in Figure 3 using both DI and saline water. The system is illuminated by a CW with the maximum amplitude allowed for safety standards [35]. In Figure 4b the temperature evolution in time is simulated assuming an excessive exposure in the order of 1 hour [32]. For the sake of clarity, the temperature rise is averaged only at the particle surface and compared with the rise occurring in the water a few centimeters away from the particle's location where the particle's influence is negligible. Note how critical is the heat released in a real solution having electrolytes concentration as in the human body. The rise quickly reaches a linear regime with time. On the other hand, the temperature in the case of DI water follows a saturated curve that is characteristic of heating governed by the Joule's effect occurring only in the particle, as it is found in optical plasmonics for noble metal nanoparticles [4,8].

To evaluate the temperature localization in and around the particle, the spatial profiles of the temperature increase are calculated as a function of distance from the particle's center for several values of time, always taken before 6 minutes (Figure 4c). A maximum irradiation time of up to 6 minutes constitutes a reasonable value [32]. Solid (dashed) lines correspond to DI (absorbing) water at 168 (555) MHz, in correspondence with the values given in Figure 4b. These profiles may result very useful to evaluate potential damages in realistic biological environments when saline solutions are used. When using DI water, the maximum temperatures occur inside the particles. The profiles for this minute-time scale resemble those for metal-nanoparticle optical heating under CW radiation in nanosecond scale [4]. In this case, the effect of the particle's heating vanishes in a few millimeters, no more than 2 cm in this case (solid lines). On the other hand, assuming that a typical monitoring study in humans could be done in a few minutes, the temperature rise holds always below 0.1 K. Note how different is the heating phenomenon using saline water to using DI-water; the temperature in water is always higher than the particle's temperature for each time value, thus giving a kind of inverted profile for the rise (dashed lines in Figure 4c). It is important to note that, in this case, the heating is not localized. Although this problem does not concern directly to RP phenomena, because it is mainly due to water heating, it is very important to evaluate realistic biomedical applications. Nonetheless, this problem can be easily avoided by using pulsed radiation or pulsed waves (PW) as well as by taking proper irradiation intervals. In this regard, the proper choice of a set-up would take into account a possible dispersion of the PW, but this is beyond the scope of this paper.

Now the study is concerned with the exploration of more complex RP particles, keeping the focus on biomedical applications. As in optical plasmonics, the shape and materials dependences of plasmonics can be used to design a specific particle. The SRPs resonances can be sensitive to the surrounding environments as well as they can give near-field confinement and/or enhancement as required. To illustrate a simple ideal example, the EM response by a spherical mm-particle having a core-shell structure is simulated, Figure 5. BS/MWCNT glass 20%wt. is selected as RF-MM for the shell and Teflon for the core since it could be easily prepared in mm-scale and safely used in biomedicine [72]. As Teflon has negligible absorption in RF, this example constitutes a way to minimize undesired heating effects since the losses occur only in the shell region, which has a shell thickness $d$ of only 0.3 mm.

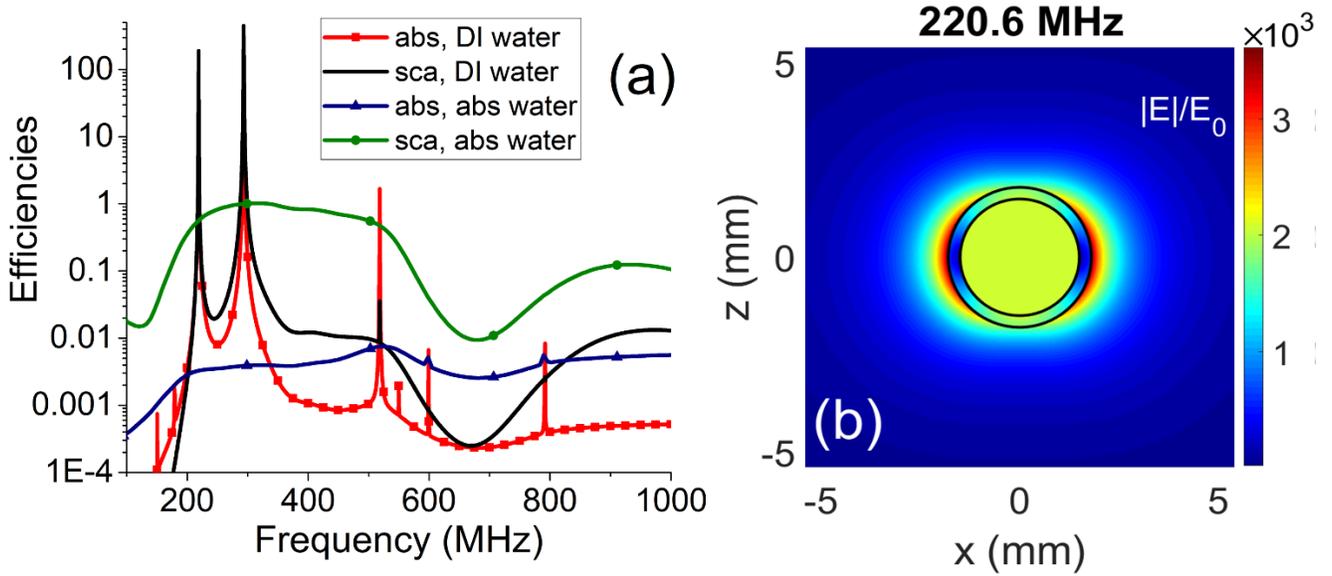

Figure 5: Electromagnetic response by a Teflon core BS/MWCNT glass 20%wt. shell particle immersed in water. The radii are 1.5 and 1.8 mm. (a) Efficiencies comparing DI vs. absorbing water with Salinity=3.2 ppt. (b) Scaled electric field around the particle for the RP resonance at 220.6 MHz when DI water is simulated. The field enhancement goes down to values seen in "usual plasmonics" when absorbing water is used (not shown).

The EM efficiencies are compared for the particle immersed in DI and absorbing water, Figure 5a. The near-field distribution of the electric field is given in Figure 5b for a particle in DI water resonating at 220.6 MHz. Note that both scattering and absorption efficiencies result very high for the case of DI water but they are strongly reduced by the effect of the water absorption, Figure 5a. As said, the screening effect of the surrounding medium prevents radiation to reach the particle and increase the resonant efficiencies due to RP excitation. In general, several multipole fields are being excited that can be appreciated more in the absorption due to DI water (red line with squares). An analysis of multipolar contributions of this case was made based on the Mie solution. Through such analysis, the two higher peaks were identified to correspond to purely dipolar SRPs. The presence of the two strong dipolar excitations in the spectra of Figure 5a is characteristic of plasmon hybridization for this structure [53,63]. The SRPs for the core-shell particle can be seen as a hybridization or combination of the respective SRPs of the bare sphere and the spherical cavity that would compose the particle [53]. This can be seen also in the map in Figure 5b for the lowest-energy SRP of the structure located at 220.6 MHz. It is clear from the map the dipolar nature of the electric-field pattern, and also that the resonance is produced by a combination of both RP excitations associated with each shell interface involved. Thus, this example illustrates that several excitations can be produced and tuned if needed by coupling several scatterers with the proper materials and shape combinations, as it is usually realized in optical plasmonics.

Another remarkable point in this particular example is the huge field enhancement in a confined space that is possible to produce with such a hybrid structure. A gigantic field enhancement of $\sim 3.5 \times 10^3$ is obtained in a space of $d = 0.3$ mm in this case. Although this value is obtained in DI water, it can still be very useful in radiofrequency applications like underwater RF identification or plasma experiments in water [73]. Applications different from biomedicine products may arise or be improved as food-treatment [38,39,74] as well as ex-vivo analysis of living samples by using DI water.

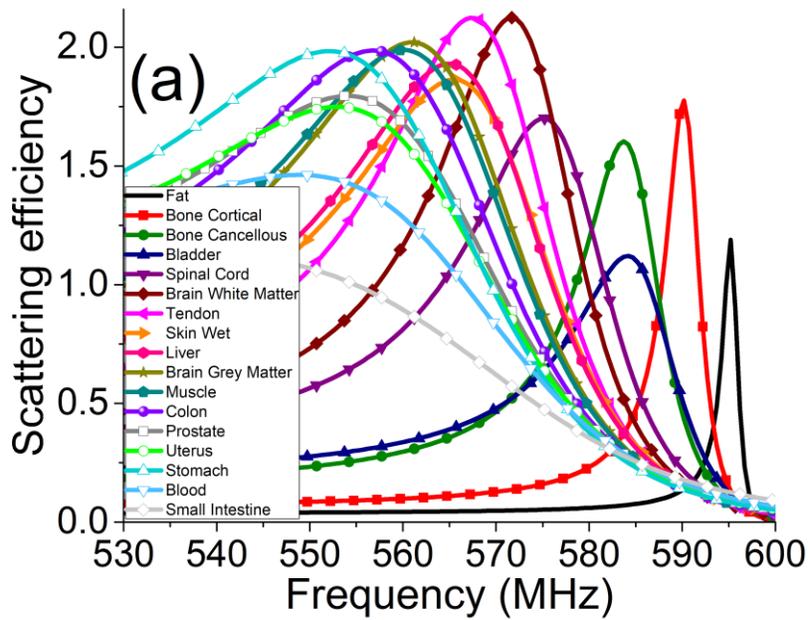

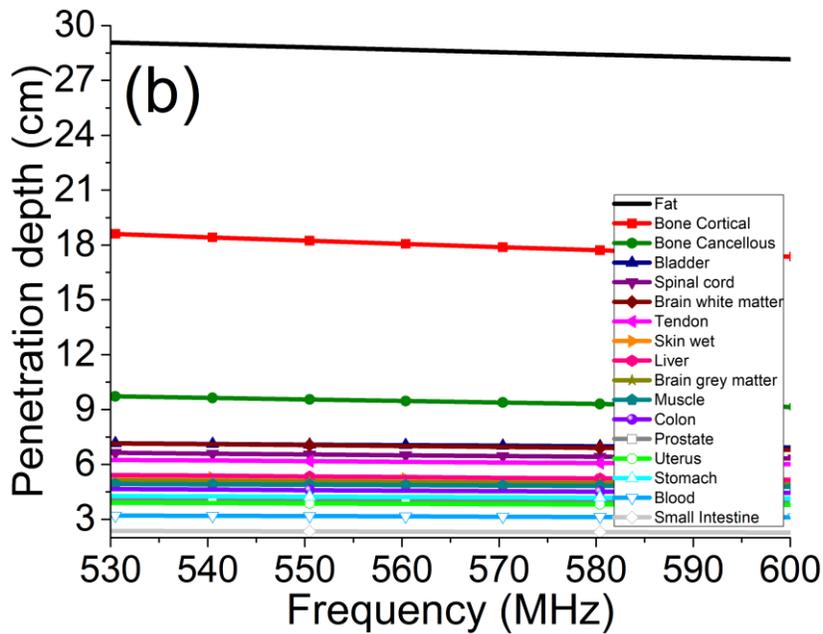

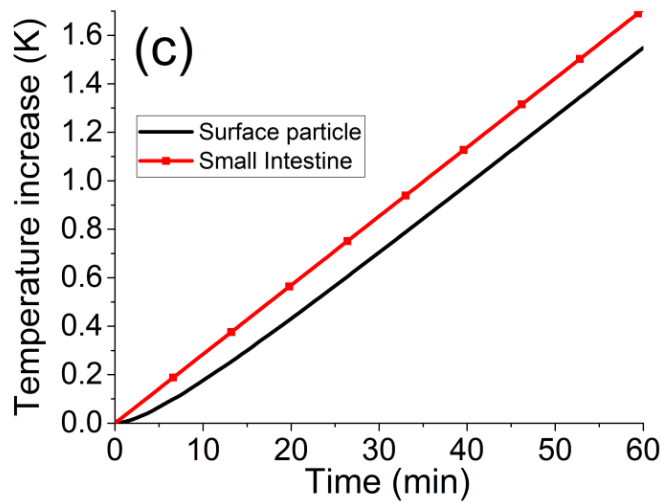

Figure 6: Applicability of bare particles of BS/MWCNT glass 20%wt. to work as monitoring antennas in several human tissues as embedding media. Size is 3.6 mm. (a) Scattering efficiency; (b) Penetration

depths. The range of frequencies is chosen to distinguish tissues by the RP frequencies. (c) Temperature-increase using the human small intestine as embedding medium at RP resonance (537.2 MHz). Data of tissues are taken from Refs. [33,75,76].

After the examination of the heat generation of solutions in a stationary regime, a system constituting a solid RP sphere immersed in human tissue is investigated, Figure 6. The tissues are considered as absorbing media by using their appropriate dielectric functions in the MHz range [33]. The EM properties of such a system are shown in Figure 6a and 6b for several tissues. Figure 6a (b) shows the scattering efficiency (penetration depth). The range of frequencies calculated is carefully chosen to allow the scatterer to have two basic features, namely to present distinguishable signals coming from each tissue (Figure 6a) and be reachable by radiation (Figure 6b). The penetration depth is defined as typical, meaning the distance for the radiation intensity to fall to 1/e (about 37%) of its original value [64]. Thus, both scattering signals and "effective" penetration will in general depend on the intensity of the incoming radiation. Remarkably, several tissues can be identified from the effective RP frequencies excited for the BS/MWCNT glass 20%wt. particle. The peaks show relatively enough quality to identify their corresponding tissues; in other words, the RP particle is highly sensitive to its bio-environment. Again, a range of frequencies is chosen such that it allows for high penetration in the tissues and appropriate wideband for information-reading purposes. An ideal experiment would take into consideration the combination of materials that results detectable in a specific organ. However, these results are found very useful as a proof of concept for RF *in-vivo* detectors.

There are other interesting frequency regions where the RP mm-antenna principle could be useful. For instance, enhanced signals are also found for each tissue around 800 MHz that can distinguish the scatterers in different tissues by the intensity and not by frequency (not shown here).

As a complement, Figure 6c shows the temperature increase of the system BS/MWCNT glass 20%wt. particle+average small intestine in the same time range and with the same incident wave as in Figure 4b. The thermal properties of the tissue are taken from [76]. This tissue was chosen because it shows the highest conductivity of all the tissues studied, thus contributing to the highest increase in temperature. As in Figure 4b, the temperature in both the particle surface and the embedding medium are evaluated. Similar to what is found in Figure 4b, the contribution of the tissue is the dominating one since its high value of losses compared with the RP particle. Although a linear trend is obtained and higher values of non-localized increase are achieved, the temperature can be controlled by the irradiation time and the type of illumination, i.e. choosing PW instead of CW. As a reference, the temperature increases up to 0.17 mK using CW in the first 6 minutes.

Finally, another interesting example is discussed. In this case, the sensitiveness that core-shell plasmonic systems usually show in optics is taken profit in the RF range to illustrate a direct human-body application. A dielectric core-RP shell particle embedded in a human small intestine is simulated for several core media. Paracetamol is used as core media in different concentrations that are represented by their corresponding dielectric functions in the RF range [77]. BS/MWCNT glass 20%wt. is chosen as the shell material. The dielectric properties of the small intestine were taken from Ref. [75]. The resultant scattering efficiencies are shown in Figure 7 as a function of frequency for several values of paracetamol concentration in molar units (M). The example consists of an illustration of the possibilities that Radioplasmonics may offer regarding drug delivery and traceability in tissues. In this example, two sets of peaks due to RP excitation are found around 600 and 800 MHz. As can be realized from Figure 7, the drug delivery could be monitored because the scattering amplitudes are different for each concentration. These concentrations mean typical values for human doses [77]. The example was conceived since there currently exist several trials to deliver and monitor drugs and/or scan in-vivo tissues with RF devices [55,78–80]. Remarkably, the success of the drug release can be estimated by comparing the spectra with the case of an empty core (violet line with down triangles in Figure 7). It is assumed that a mechanism could exist in a smart pill or device that does not allow body fluids or small tissue elements to enter the scatterer, maybe exerting pressure on the drug to push it out [80,81]. In this way, the core will be empty and the scattering is exclusively ruled by the RP shell.

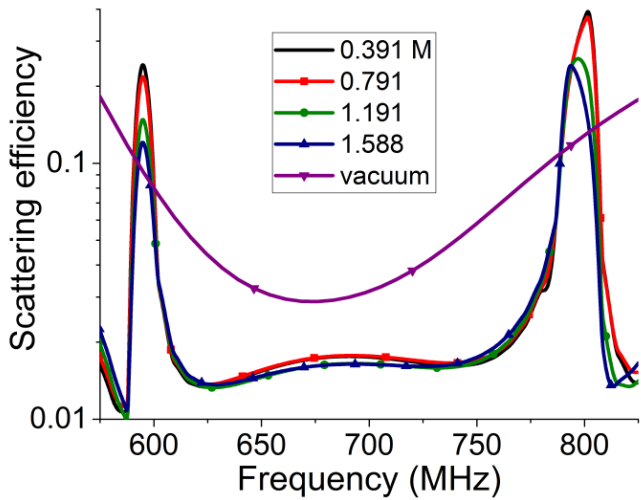

Figure 7: Scattering efficiencies for a paracetamol core- BS/MWCNT glass 20%wt. shell particle immersed in a human small intestine for several core media. The spectra simulate the use of different paracetamol concentrations in the particle core (see molarities in the inset) as an illustration of possible *in vivo* and real-time monitoring of drug delivery in humans. Line in violet with down triangles corresponds to an empty core or vacuum, simulating the total delivery of the drug in the tissue.

It is not the purpose of this work to design precise RF devices for different applications but to show the tremendous possibilities that excitation of RP devices may allow in different host media. Specifically, several drugs can be used if their permittivity functions are known as insulin in diseases associated with diabetes [78,80]. Besides, this kind of device could be applied in other water applications like water desalinization or underwater communication.

**CONCLUSION**

Radioplasmonics consists of an extension of optical plasmonics to the radiofrequency range, based on realistic metamaterials able to mimic metal conduction in MHz to GHz regions. It naturally generalizes and complements the recently-developed techniques based on "spoof" plasmons. Other novel concepts as Terrestrial surface plasmon", and "Soft-Plasmonics" are automatically included in the discussion. This work defined for the first time plasmonics of composite systems on a human-living scale. The perspectives that bring the concept of Radioplasmonics were fully explored. Different from a recent work by the author, in which the radioplasmonic absorption of microparticles was taken to enhance photoacoustic signals, the full electromagnetic properties of radioplasmonic particles were taken here into consideration, namely absorption and scattering produced by bigger particles. Centimeter or millimeter particles of several shapes could be easily synthesizable with the current technologies. The near-fields of these scatterers revealed that huge field enhancements and confinements can be achieved.

As a consequence of the present studies, easy and low-cost radiofrequency experiments could be designed to study complex plasmonic phenomena and to trace similarities with smaller-scale (optical) systems. Scatterers in air were proven to be useful for antenna design. Specifically, the new sub-wavelength antennas turned to be more efficient than conventional antennas and pave the way to build miniaturized structures for any modern telecommunication devices. Every future design on radiofrequency networks/ radiofrequency identification/ near-field communication should take into consideration the novel plasmonic phenomena stated here.

An extensive study of simple scatterers in water-like and biological media was also performed. Furthermore, by using a few illustrative examples based on Radioplasmonics, ideas were given to improve remarkable biomedical applications as real-time, *in-vivo* monitoring drug-delivery in real human body tissues. As a complement, the RF heating and scattering in absorbing tissues/ saline solutions were studied in the worst realistic scenario, consisting of using continuous illumination and maximum allowed power. As absorbing media as saline water was used, the results may find applicability in technologies like water desalinization, radiofrequency ablation, underwater communication, and food treatment, among others.

## AUTHOR INFORMATION

Corresponding Author

* Email: mabraham@exa.unicen.edu.ar

Present Addresses

† *Istituto Italiano di Tecnologia, Via Morego 30 16163 Genova, Italy.*

‡ Instituto de Física Arroyo Seco, IFAS (UNCPBA), Tandil, Argentina.

¶ CIFICEN (UNCPBA-CICPBA-CONICET), Grupo de Plasmas Densos, Pinto 399, 7000 Tandil, Argentina.

## ACKNOWLEDGMENT

RMAE would like to thank Dr. Cristian D'Angelo from IFAS-UNCPBA and Dr. Francesco De Angelis from IIT for stimulating discussions on the topic. The author also thanks Plasmon Nanotechnologies group for their hospitality in IIT and the permission to use their licensed software.

## ABBREVIATIONS

EM, electromagnetic; MM/s, metamaterial/s; RF, radiofrequency; RFID, Radiofrequency Identification; NFC, Near-field Communication; SPs, spoof plasmons; RP, radioplasmonic; GHz, gigahertz; MHz, megahertz; CW, continuous wave; PDMS, Polydimethylsiloxane; APu/Al, Acrylic/Polyurethane/Aluminum; MWCNT, multi-walled carbon nanotubes; CNF, carbon nanofiber; NiO, Nickel Oxide; TiN, titanium nitride; PPy, Polypyrrole; CNTs, carbon nanotubes; SI, Supplementary Information; SRPs, surface radioplasmons; DI, distilled; BS, composition of $Bi_2O_3$ and $SiO_2$.

## FUNDING

This research was supported by Facultad de Ciencias Exactas, Universidad Nacional del Centro de la Provincia de Buenos Aires from Argentina and Istituto Italiano di Tecnologia from Italy.

## CONFLICTS OF INTEREST/ COMPETING INTERESTS

The author has no conflicts of interest/competing interests to disclose.

## AUTHOR CONTRIBUTIONS

R.M.A.-E. conceived the idea, made of the calculations, analyzed the results, and wrote the manuscript.

## ETHICS APPROVAL

Not Applicable

# REFERENCES

[1] Garoli D, Calandrini E, Giovannini G, Hubarevich A, Caligiuri V and Angelis F D 2019 Nanoporous gold metamaterials for high sensitivity plasmonic sensing *Nanoscale Horiz.* **4** 1153–7

[2] Novotny L and Hecht B 2006 *Principles of Nano-Optics* (Cambridge: Cambridge University Press)

[3] Li M, Cushing S K and Wu N 2015 Plasmon-Enhanced Optical Sensors: A Review *Analyst* **140** 386–406

[4] Baffou G 2017 *Thermoplasmonics: Heating Metal Nanoparticles Using Light* (Cambridge: Cambridge University Press)

[5] Zhang X, Zheng Y, Liu X, Lu W, Dai J, Lei D Y and MacFarlane D R 2015 Hierarchical Porous Plasmonic Metamaterials for Reproducible Ultrasensitive Surface-Enhanced Raman Spectroscopy *Advanced Materials* **27** 1090–6

[6] Nam J-M, Oh J-W, Lee H and Suh Y D 2016 Plasmonic Nanogap-Enhanced Raman Scattering with Nanoparticles

[7] Messina G C, Malerba M, Zilio P, Miele E, Dipalo M, Ferrara L and De Angelis F 2015 Hollow plasmonic antennas for broadband SERS spectroscopy *Beilstein J Nanotechnol* **6** 492–8

[8] Giannini V, Fernández-Domínguez A I, Heck S C and Maier S A 2011 Plasmonic Nanoantennas: Fundamentals and Their Use in Controlling the Radiative Properties of Nanoemitters *Chem. Rev.* **111** 3888–912

[9] Rodríguez-Oliveros R, Sánchez-Gil J A, Giannini V, Macías D and Rivas J G Plasmon Optical Nanoantennas: Characterization, Design, and Applications in Nanophotonics

[10] Nazari M, Xi M, Lerch S, Alizadeh M H, Ettinger C, Akiyama H, Gillespie C, Gummuluru S, Erramilli S and Reinhard B M 2017 Plasmonic Enhancement of Selective Photonic Virus Inactivation *Scientific Reports* **7** 11951

[11] A. Goerlitzer E S, E. Speichermann L, A. Mirza T, Mohammadi R and Vogel N 2020 Addressing the plasmonic hotspot region by site-specific functionalization of nanostructures *Nanoscale Advances* **2** 394–400

[12] Urbas A M, Jacob Z, Negro L D, Engheta N, Boardman A D, Egan P, Khanikaev A B, Menon V, Ferrera M, Kinsey N, DeVault C, Kim J, Shalaev V, Boltasseva A, Valentine J, Pfeiffer C, Grbic A, Narimanov E, Zhu L, Fan S, Alù A, Poutrina E, Litchinitser N M, Noginov M A, MacDonald K F, Plum E, Liu X, Nealey P F, Kagan C R, Murray C B, Pawlak D A, Smolyaninov I I, Smolyaninova V N and Chanda D 2016 Roadmap on optical metamaterials *J. Opt.* **18** 093005

[13] Chen D and Zheng X 2018 Multi-material Additive Manufacturing of Metamaterials with Giant, Tailorable Negative Poisson ' s Ratios *Scientific Reports* **8** 9139

[14] Pendry J B, Martín-Moreno L and Garcia-Vidal F J 2004 Mimicking Surface Plasmons with Structured Surfaces *Science* **305** 847–8

[15] Pendry J B, Schurig D and Smith D R 2006 Controlling Electromagnetic Fields *Science* **312** 1780–2

[16] Klotz G, Malléjac N, Guenneau S and Enoch S 2019 Controlling frequency dispersion in electromagnetic invisibility cloaks *Scientific Reports* **9** 6022

[17] Bonache J, Zamora G, Paredes F, Zuffanelli S, Aguilà P and Martín F 2016 Controlling the Electromagnetic Field Confinement with Metamaterials *Scientific Reports* **6** 37739

[18] Stutzman W L and Thiele G A 2012 *Antenna Theory and Design* (John Wiley & Sons)

[19] Agio M and Alù A 2013 *Optical Antennas* (Cambridge: Cambridge University Press)

[20] Lazaro A, Villarino R and Girbau D 2018 A Survey of NFC Sensors Based on Energy Harvesting for IoT Applications *Sensors* **18** 3746

[21] Qin F, Zhang Q and Xiao J-J 2016 Sub-wavelength Unidirectional Antenna Realized by Stacked Spoof Localized Surface Plasmon Resonators *Scientific Reports* **6** 29773

[22] Zhang Y and Han Z 2015 Spoof surface plasmon based planar antennas for the realization of Terahertz hotspots *Scientific Reports* **5** 18606

[23] Tang W X, Zhang H C, Ma H F, Jiang W X and Cui T J 2019 Concept, Theory, Design, and Applications of Spoof Surface Plasmon Polaritons at Microwave Frequencies *Advanced Optical Materials* **7** 1800421

[24] Liao Z, Luo G Q, Wu X Y, Cai B G, Cao Pan B and Pan Y J 2020 A Horizontally Polarized Omnidirectional Antenna Based on Spoof Surface Plasmons *Front. Phys.* **8**

[25] Garcia-Vidal F J, Martín-Moreno L and Pendry J B 2005 Surfaces with holes in them: new plasmonic metamaterials *J. Opt. A: Pure Appl. Opt.* **7** S97–101

[26] Shao R L, Li B, Yang L and Zhou Y J 2019 Electrically small multiband antenna based on spoof localized surface plasmons ed J Zhao, W Tang and S Hongyu *EPJ Appl. Metamat.* **6** 11

[27] Dai L H, Tan C and Zhou Y J 2020 Ultrawideband Low-Profile and Miniaturized Spoof Plasmonic Vivaldi Antenna for Base Station *Applied Sciences* **10** 2429

[28] Maksymov I S, Staude I, Miroshnichenko A E and Kivshar Y S 2012 Optical Yagi-Uda nanoantennas *Nanophotonics* **1** 65–81

[29] Rocca R L, Messina G C, Dipalo M, Shalabaeva V and Angelis F D 2015 Out-of-Plane Plasmonic Antennas for Raman Analysis in Living Cells *Small* **11** 4632–7

[30] Huang X and El-Sayed M A 2011 Plasmonic photo-thermal therapy (PPTT) *Alexandria Journal of Medicine* **47** 1–9

[31] Lim W Q and Gao Z 2016 Plasmonic nanoparticles in biomedicine *Nano Today* **11** 168–88

[32] Boehm T, Malich A, Goldberg S N, Reichenbach J R, Hilger I, Hauff P, Reinhardt M, Fleck M and Kaiser W A 2002 Radio-frequency Tumor Ablation: Internally Cooled Electrode versus Saline-enhanced Technique in an Aggressive Rabbit Tumor Model *Radiology* **222** 805–13

[33] Gabriel C, Gabriel S and Corthout E 1996 The dielectric properties of biological tissues: I. Literature survey *Phys. Med. Biol.* **41** 2231–49

[34] Abraham-Ekeroth R M and De Angelis F Radioplasmonics: plasmonic transducers in the radiofrequency regime for resonant thermo-acoustic imaging in deep tissues. *ACS Photonics (accepted manuscript)*

[35]     Anon 2019 IEEE Standard for Safety Levels with Respect to Human Exposure to Electric, Magnetic, and Electromagnetic Fields, 0 Hz to 300 GHz *IEEE Std C95.1-2019 (Revision of IEEE Std C95.1-2005/ Incorporates IEEE Std C95.1-2019/Cor 1-2019)* 1–312

[36]     Tamarov K, Gongalsky M, Osminkina L, Huang Y, Omar M, Yakunin V, Ntziachristos V, Razansky D and Timoshenko V 2017 Electrolytic conductivity-related radiofrequency heating of aqueous suspensions of nanoparticles for biomedicine *Phys. Chem. Chem. Phys.* **19** 11510–7

[37]     Nicoletti G, Cornaglia A I, Faga A and Scevola S 2014 The Biological Effects of Quadripolar Radiofrequency Sequential Application: A Human Experimental Study *Photomed Laser Surg* **32** 561–73

[38]     Wang Y, Wig T D, Tang J and Hallberg L M 2003 Sterilization of Foodstuffs Using Radio Frequency Heating *Journal of Food Science* **68** 539–44

[39]     Geveke D J, Brunkhorst C and Fan X 2007 Radio frequency electric fields processing of orange juice *Innovative Food Science & Emerging Technologies* **8** 549–54

[40]     Hou Q, Yan K, Fan R, Zhang Z, Chen M, Sun K and Cheng C 2015 Experimental realization of tunable negative permittivity in percolative Fe78Si9B13/epoxy composites *RSC Adv.* **5** 9472–5

[41]     Shi Z, Mao F, Wang J, Fan R and Wang X 2015 Percolative silver/alumina composites with radio frequency dielectric resonance-induced negative permittivity *RSC Adv.* **5** 107307–12

[42]     Tallman T N 2020 The effect of thermal loading on negative permittivity in carbon nanofiber/silicone metacomposites *Materials Today Communications* **22** 100843

[43]     Cheng C, Fan R, Ren Y, Ding T, Qian L, Guo J, Li X, An L, Lei Y, Yin Y and Guo Z 2017 Radio frequency negative permittivity in random carbon nanotubes/alumina nanocomposites *Nanoscale* **9** 5779–87

[44]     Qu Y, Du Y, Fan G, Xin J, Liu Y, Xie P, You S, Zhang Z, Sun K and Fan R 2019 Low-temperature sintering Graphene/CaCu3Ti4O12 nanocomposites with tunable negative permittivity *Journal of Alloys and Compounds* **771** 699–710

[45]     Estevez D, Qin F, Luo Y, Quan L, Mai Y-W, Panina L and Peng H-X 2019 Tunable negative permittivity in nano-carbon coated magnetic microwire polymer metacomposites *Composites Science and Technology* **171** 206–17

[46]     Sun K, Dong J, Wang Z, Wang Z, Fan G, Hou Q, An L, Dong M, Fan R and Guo Z 2019 Tunable Negative Permittivity in Flexible Graphene/PDMS Metacomposites *J. Phys. Chem. C* **123** 23635–42

[47]     Wang Z, Sun K, Xie P, Liu Y, Gu Q and Fan R 2020 Permittivity transition from positive to negative in acrylic polyurethane-aluminum composites *Composites Science and Technology* **188** 107969

[48]     Fujii M 2016 A new mode of radio wave diffraction via the terrestrial surface plasmon on mountain range *Radio Science* **51** 1396–412

[49]     Fujii M 2013 Theory of ground surface plasma wave associated with pre-earthquake electrical charges *Radio Science* **48** 122–30

[50]     Jacak W A 2016 Plasmons in Finite Spherical Electrolyte Systems: RPA Effective Jellium Model for Ionic Plasma Excitations *Plasmonics* **11** 637–51

[51]     Jacak J and Jacak W 2019 Plasmons and Plasmon–Polaritons in Finite Ionic Systems: Toward Soft-Plasmonics of Confined Electrolyte Structures *Applied Sciences* **9** 1159

[52]	David C 2018 Plasmonic Properties of Electrolytes Beyond Classical Nanophotonics — A Two-Fluid, Hydrodynamic Approach to Nonlocal Soft Plasmonics *2018 Progress in Electromagnetics Research Symposium (PIERS-Toyama)* 2018 Progress in Electromagnetics Research Symposium (PIERS-Toyama) pp 490–5

[53]	Prodan E and Nordlander P 2004 Plasmon hybridization in spherical nanoparticles *The Journal of Chemical Physics* **120** 5444–54

[54]	Cai X, Zhao S, Hu M, Xiao J, Zhang N and Yang J 2017 Water based fluidic radio frequency metamaterials *Journal of Applied Physics* **122** 184101

[55]	Ma Y, Luo Z, Steiger C, Traverso G and Adib F 2018 Enabling deep-tissue networking for miniature medical devices *Proceedings of the 2018 Conference of the ACM Special Interest Group on Data Communication* SIGCOMM '18 (New York, NY, USA: Association for Computing Machinery) pp 417–31

[56]	Fu Q and Sun W 2001 Mie theory for light scattering by a spherical particle in an absorbing medium *Appl. Opt., AO* **40** 1354–61

[57]	Sudiarta I W and Chylek P 2001 Mie-scattering formalism for spherical particles embedded in an absorbing medium *J. Opt. Soc. Am. A, JOSAA* **18** 1275–8

[58]	Bohren C F and Huffman D R 1998 *Absorption and Scattering of Light by Small Particles* (Weinheim: Wiley-VCH)

[59]	Suzuki H and Lee I-Y S 2008 Calculation of the Mie scattering field inside and outside a coated spherical particle *International Journal of Physical Sciences* **3** 038–41

[60]	Qu Y, Li Y, Xu C, Fan G, Xie P, Wang Z, Liu Y, Wu Y and Fan R 2018 Metacomposites: functional design via titanium nitride/nickel(II) oxide composites towards tailorable negative dielectric properties at radio-frequency range *J Mater Sci: Mater Electron* **29** 5853–61

[61]	Xu C, Qu Y, Fan G, Ren H, Chen J, Liu Y, Wu Y and Fan R 2018 Low loading carbon nanotubes supported polypyrrole nano metacomposites with tailorable negative permittivity in radio frequency range *Organic Electronics* **63** 362–8

[62]	Novotny L 2007 Effective Wavelength Scaling for Optical Antennas *Phys. Rev. Lett.* **98** 266802

[63]	Zhang J, Liao Z, Luo Y, Shen X, Maier S A and Cui T J 2017 Spoof plasmon hybridization *Laser & Photonics Reviews* **11** 1600191

[64]	Jackson J D 1999 *Classical electrodynamics* (New York : Wiley)

[65]	Stogryn A 1971 Equations for Calculating the Dielectric Constant of Saline Water (Correspondence) *IEEE Transactions on Microwave Theory and Techniques* **19** 733–6

[66]	Klein L and Swift C 1977 An improved model for the dielectric constant of sea water at microwave frequencies *IEEE Transactions on Antennas and Propagation* **25** 104–11

[67]	Somaraju R and Trumpf J 2006 Frequency, Temperature and Salinity Variation of the Permittivity of Seawater *IEEE Transactions on Antennas and Propagation* **54** 3441–8

[68]	Haldar T, Kumar U, Yadav B C and Kumar V V R K 2020 Tunable negative permittivity of Bi2O3–SiO2/MWCNT glass-nanocomposites at radio frequency region *J Mater Sci: Mater Electron* **31** 11791–800

[69]   Sobotka L, Allison S and Stanga Z 2008 Basics in clinical nutrition: Water and electrolytes in health and disease *e-SPEN, the European e-Journal of Clinical Nutrition and Metabolism* **3** e259–66

[70]   Tamarov K P, Osminkina L A, Zinovyev S V, Maximova K A, Kargina J V, Gongalsky M B, Ryabchikov Y, Al-Kattan A, Sviridov A P, Sentis M, Ivanov A V, Nikiforov V N, Kabashin A V and Timoshenko V Y 2014 Radio frequency radiation-induced hyperthermia using Si nanoparticle-based sensitizers for mild cancer therapy *Scientific Reports* **4** 7034

[71]   Liu X, Chen H, Chen X, Alfadhl Y, Yu J and Wen D 2015 Radiofrequency heating of nanomaterials for cancer treatment: Progress, controversies, and future development *Applied Physics Reviews* **2** 011103

[72]   Naftalovich R, Naftalovich D and Greenway F 2016 Polytetrafluoroethylene Ingestion as a Way to Increase Food Volume and Hence Satiety Without Increasing Calorie Content *Journal of diabetes science and technology*

[73]   Maehara T, Toyota H, Kuramoto M, Iwamae A, Tadokoro A, Mukasa S, Yamashita H, Kawashima A and Nomura S 2006 Radio Frequency Plasma in Water *Jpn. J. Appl. Phys.* **45** 8864

[74]   Geveke D J 2005 12 - Non-thermal Processing By Radio Frequency Electric Fields *Emerging Technologies for Food Processing* ed D-W Sun (London: Academic Press) pp 307–22

[75]   Andreuccetti D, Fossi R and Petrucci C 1997 An Internet resource for the calculation of the dielectric properties of body tissues in the frequency range 10 Hz - 100 GHz, IFAC-CNR, Florence (Italy), 1997. Based on data published by C.Gabriel et al. in 1996. [Online]. *An Internet resource for the calculation of the dielectric properties of body tissues in the frequency range 10 Hz - 100 GHz*

[76]   Hasgall P A, F. Di Gennaro, Baumgartner C, Neufeld E, Lloyd B, Gosselin M C, Payne D, Klingenböck A and Kuster N 2018 IT'IS Database for thermal and electromagnetic parameters of biological tissues - Version 4.0 itis.swiss/database

[77]   Cheng E M, Fareq M, Abdullah F S, Wee F H, Khor S F, Lee Y S, Afendi M, Shukry M, Shahriman A B, Khairunizanm W and Liyana Z Y 2013 Dielectric spectroscopy of pharmaceutical drug (Paracetamol) dosage in water *2013 IEEE International RF and Microwave Conference (RFM)* 2013 IEEE International RF and Microwave Conference (RFM) pp 409–13

[78]   Juan C G, Bronchalo E, Potelon B, Quendo C and Sabater-Navarro J M 2019 Glucose Concentration Measurement in Human Blood Plasma Solutions with Microwave Sensors *Sensors (Basel)* **19**

[79]   Abid A, O'Brien J M, Bensel T, Cleveland C, Booth L, Smith B R, Langer R and Traverso G 2017 Wireless Power Transfer to Millimeter-Sized Gastrointestinal Electronics Validated in a Swine Model *Scientific Reports* **7** 46745

[80]   Abramson A, Caffarel-Salvador E, Khang M, Dellal D, Silverstein D, Gao Y, Frederiksen M R, Vegge A, Hubálek F, Water J J, Friderichsen A V, Fels J, Kirk R K, Cleveland C, Collins J, Tamang S, Hayward A, Landh T, Buckley S T, Roxhed N, Rahbek U, Langer R and Traverso G 2019 An ingestible self-orienting system for oral delivery of macromolecules *Science* **363** 611–5

[81]   Kong Y L, Zou X, McCandler C A, Kirtane A R, Ning S, Zhou J, Abid A, Jafari M, Rogner J, Minahan D, Collins J E, McDonnell S, Cleveland C, Bensel T, Tamang S, Arrick G, Gimbel A, Hua T, Ghosh U, Soares V, Wang N, Wahane A, Hayward A, Zhang S, Smith B R, Langer R and Traverso G 2019 3D-Printed Gastric Resident Electronics *Adv Mater Technol* **4**